\documentclass[a4paper,aps,prd,twocolumn,10pt,superscriptaddress,nofootinbib,showpacs]{revtex4-1}
\usepackage{graphicx,amsmath,amsfonts,amssymb,revsymb,dcolumn,epsfig,bm,xspace}
\usepackage[hyperfootnotes=false]{hyperref}
\usepackage[latin1]{inputenc}
\usepackage{color}

\newcommand{\lie}{\pounds}
\newcommand{\del}{\partial}

\newcommand{\indspace}{\quad}
\DeclareFontFamily{U}{mathx}{\hyphenchar\font45}
\DeclareFontShape{U}{mathx}{m}{n}{<-> mathx10}{}
\DeclareSymbolFont{mathx}{U}{mathx}{m}{n}
\DeclareMathAccent{\widebar}{0}{mathx}{"73}
\newcommand{\backdec}[1]{\bar{#1}}
\newcommand{\wbackdec}[1]{\widebar{#1}}
\newcommand{\g}{g}

\newcommand{\bg}{{\backdec{g}}}
\newcommand{\n}{v}
\newcommand{\bn}{{\backdec{\n}}}
\newcommand{\dn}{{\delta\n}}
\newcommand{\sn}{\mathsf{v}}
\newcommand{\h}{\gamma}
\newcommand{\bh}{\backdec{\h}}

\newcommand{\hp}[1]{\h\left[#1\right]}
\newcommand{\bhp}[1]{\bh\left[#1\right]}
\newcommand{\cd}{\nabla}
\newcommand{\bcd}{\wbackdec{\nabla}}
\newcommand{\scd}{D}
\newcommand{\bscd}{\wbackdec{\scd}}
\newcommand{\scp}{\parallel}
\newcommand{\blap}{\wbackdec{\scd}^2}
\newcommand{\blapK}{\blap_{\K}}

\newcommand{\G}{G}

\newcommand{\kp}{\kappa}

\newcommand{\ddg}{\xi}

\newcommand{\A}{\phi}
\newcommand{\B}{B}
\newcommand{\C}{C}
\newcommand{\CS}{\psi}
\newcommand{\CSI}{\Psi}
\newcommand{\CSvms}{\mathcal{U}}
\newcommand{\MS}{\zeta}

\newcommand{\CTD}{W}
\newcommand{\BS}{\mathcal{\B}}
\newcommand{\BV}{\mathtt{\B}}
\newcommand{\CSD}{\mathcal{E}}
\newcommand{\CV}{\mathtt{F}}

\newcommand{\EC}{\mathcal{K}}
\newcommand{\bEC}{\wbackdec{\EC}}

\newcommand{\EX}{\Theta}
\newcommand{\CEX}{Z}
\newcommand{\dEX}{{\delta\EX}}
\newcommand{\dEXf}{\fdelta\EX}

\newcommand{\bEX}{\wbackdec{\EX}}
\newcommand{\SH}{\sigma}
\newcommand{\dSH}{\delta\SH}
\newcommand{\dSHs}{\dSH^{\text{s}}{}}
\newcommand{\dSHv}{\dSH^{\text{v}}{}}
\newcommand{\dSHsf}{\fdelta\SH^{\text{s}}{}}

\newcommand{\SR}{\mathcal{R}}
\newcommand{\RR}{\mathsf{R}}
\newcommand{\dSR}{{\delta\SR}}
\newcommand{\dSRf}{\fdelta\SR}
\newcommand{\bSR}{\wbackdec{\SR}}
\newcommand{\K}{K}
\newcommand{\bK}{\wbackdec{\K}}
\newcommand{\ac}{a}

\newcommand{\dac}{{\delta\ac}}
\newcommand{\dacf}{{\fdelta\ac}}

\newcommand{\T}{T}

\newcommand{\mf}{\varphi}
\newcommand{\dmf}{\delta\mf}

\newcommand{\vms}{\mathcal{V}}
\newcommand{\bmf}{\backdec{\mf}}

\newcommand{\fdelta}{\delta^\text{f}}

\newcommand{\p}{p}
\newcommand{\Cp}{Y}

\newcommand{\ed}{\rho}
\newcommand{\Ced}{X}

\newcommand{\cs}{c_s}
\newcommand{\bcs}{\backdec{c}_s}

\newcommand{\bp}{\backdec{p}}
\newcommand{\bed}{\backdec{\ed}}

\newcommand{\w}{w}
\newcommand{\bw}{\backdec{\w}}
\newcommand{\x}{x}

\newcommand{\kcond}{\varepsilon}
\newcommand{\kcondac}{\tilde{\kcond}}
\newcommand{\mode}{Q}
\newcommand{\mki}[1]{k_{#1}}
\newcommand{\mkco}{k_{i_\text{max}}}
\newcommand{\mkcoz}{k_{i_\text{max}0}}
\newcommand{\dmkcoz}{\dot{k}_{i_\text{max}0}}
\newcommand{\lp}{l_\text{p}}
\newcommand{\HR}{R_H}

\begin{document}
\title{Comments on ``Growth of Covariant Perturbations in the Contracting Phase of a Bouncing Universe'' by A.Kumar}

\author{N.~Pinto-Neto} \email{nelson.pinto@pq.cnpq.br}
\author{S.~D.~P.~Vitenti} \email{vitenti@cbpf.br}

\affiliation{Centro Brasileiro de Pesquisas F\'{\i}sicas,
Rua Dr.\ Xavier Sigaud 150 \\
22290-180, Rio de Janeiro -- RJ, Brasil}

\date{\today}

\begin{abstract}

A recent paper by Kumar (2012) (hereafter K12) claimed that in a contracting model, described by perturbations around a collapsing Friedmann model containing dust or radiation, the perturbations can grow in such a way that the linearity conditions would become invalid. This conclusion is not correct due to the following facts: first, it is claimed that the linearity conditions are not satisfied, but nowhere in K12 the amplitudes of the perturbations were in fact estimated. Therefore, without such estimates, the only possible conclusion from this work is the well known fact that the perturbations indeed grow during contraction, which, per se, does not imply that the linearity conditions become invalid. Second, some evaluations of the linearity conditions are incorrect because third other terms, instead of the appropriate second order ones, are mistakenly compared with first order terms, yielding artificially fast growing conditions. Finally, it is claimed that the results of K12 are in sharp contrast with the results of the paper by Vitenti and Pinto-Neto (2012) (hereafter VPN12), because the former was obtained in a gauge invariant way. However, the author of K12 did not realized that the evolution of the perturbations were also calculated in a gauge invariant way in VPN12, but some of the linearity conditions which are necessary to be checked cannot be expressed in terms of gauge invariant quantities. In the present work, the incorrect or incomplete statements of K12 are clarified and completed, and it is shown that all other correct results of K12 were already present in VPN12, whose conclusions remain untouched, namely, that cosmological perturbations of quantum mechanical origin in a bouncing model can remain in the linear regime all along the contracting phase and at the bounce itself for a wide interval of energy scales of the bounce, ranging from the nucleosynthesis energy scale up to some few orders of magnitude below the Planck energy.

\end{abstract}


\maketitle

\section{Introduction}

We comment on the paper by Kumar 2012 recently published in~\cite{Kumar2012}. This paper discusses the evolution of perturbations around a Friedmann model during a contracting phase, using the covariant gauge invariant approach described in~\cite{Ellis1989}. From gauge invariant equations of motion, the author extracted the linearity conditions by comparing the second order terms (in most of the examples) with the first order ones, in the cases of cosmological models with matter content described by dust and radiation. He then shows that the second order terms grow faster than the first order ones, and from this calculation alone he seems to conclude that near the bounce the linear regime becomes invalid. More specifically, in the abstract of K12 it is said that the linearity conditions are not satisfied in the bounce phase, while in the section V and in the Conclusion of this paper a softer statement is made, namely, that the linearity conditions {\bf may} be invalidated at this phase. While the former claim would be a new, important but erroneous conclusion, the later is a general remark already present in virtually any paper considering contracting phases in Friedmann models. In this comment we will show in detail why the former conclusion cannot be sustained, and that the results presented in our work VPN12~\cite{Vitenti2012} are strictly correct.

\section{Linearity Conditions}

In our paper VPN12, we have obtained a full set of linearity conditions for the evolution of cosmological perturbations in a Friedmann background space-time. This set have been obtained through the analysis of the equations of motion of all relevant geometrical objects and their perturbative expansions, and by comparing the background with first order terms. We have also shown, however, that the perturbative hypothesis must be checked in other fundamental places, like in the expansion of the physical metric $g^{\mu\nu}$ in terms of the perturbations $\delta g_{\mu\nu}$. For such quantities, it is not possible to obtain a gauge invariant condition. Therefore, even if one can combine many linearity conditions in order to express them directly in terms of gauge invariant variables, one still has to show that there is at least one gauge where the remaining conditions are valid.

Even though the conditions obtained in K12 were calculated in the covariant formalism, it is easy to translate such conditions to the usual gauge invariant formalism based on the Bardeen's gauge invariant variables~\cite{Bardeen1980}. In order to do so, we constructed a map between the so called covariant perturbation variables and the Bardeen variables. In order to facilitate the comparison, we used the notation of~\cite{Mukhanov1992} for the Bardeen gauge invariant variables, and we took from~\cite{Vitenti2013} the expressions relating the metric perturbations and the other perturbed tensors.

\subsection{General Expressions}

In this section we will summarize some results from~\cite{Vitenti2013} which are necessary to perform the comparison with the results from K12. We take a Lorentizian metric $\g_{\mu\nu}$ with signature $(-1,1,1,1)$ to represent the physical metric, and we define
\begin{equation}
\ddg_{\mu\nu} \equiv \g_{\mu\nu} - \bg_{\mu\nu} = 2\A\bn_\mu\bn_\nu+2\B_{(\mu}\bn_{\nu)} + 2\C_{\mu\nu},
\end{equation}
where
\begin{align*}
\A \equiv \frac12 \ddg_{\mu\nu}\bn^\mu\bn^\nu, \quad \B^\mu \equiv - \bhp{\ddg_{\nu}{}^{\mu}\bn^\nu},\quad \C_{\mu\nu} \equiv \frac{\bhp{\ddg_{\mu\nu}}}{2},
\end{align*}
$\bn^\mu$ is the normal vector field ($\bn^\mu\bn_\mu = -1$) representing the Friedmann's isotropic observers, $\bh_{\mu\nu} = \bg_{\mu\nu} + \bn_\mu\bn_\nu$ is the projector on the background space-like hyper-surfaces, and $\bhp{T_{\mu_1\dots}^{\quad\nu_1\dots}} = \bh_{\mu_1}{}^{\alpha_1}\dots\bh_{\beta_1}{}^{\nu_1}\dots{}T_{\alpha_1\dots}^{\quad\beta_1\dots}$. The over-bar designates the background variables. The physical tensors will have their indices raised and lowered by $\g_{\mu\nu}$, and the background ones by $\bg_{\mu\nu}$. Using the scalar, vector and tensor decomposition (see~\cite{Stewart1990}) we rewrite the metric perturbations as
\begin{align*}
\B_\mu &= \BS_{\scp\mu} + \BV_\mu, \\
\C_{\mu\nu} &= \CS\h_{\mu\nu} - \CSD_{\scp\mu\nu} + \CV_{(\nu\scp\mu)} + \CTD_{\mu\nu},
\end{align*}
where $\BV^\mu{}_{\scp\mu} = \CV^\mu{}_{\scp\mu} = \CTD_\mu{}^\nu{}_{\scp\nu} = \CTD_\mu{}^\mu = 0$.

For a global foliation described by a normal vector field $\n^\mu$ ($\n^\mu\n_\mu = -1$) we have
\begin{equation}\label{eq:def:EC:2}
\cd_\mu\n_\nu = \EC_{\mu\nu} - \n_\mu\ac_\nu,
\end{equation}
where $\cd_\mu$ is the covariant derivative compatible with $\g_{\mu\nu}$, i.e., $\cd_\alpha\g_{\mu\nu} = 0$, $\EC_{\mu\nu}$ is the extrinsic curvature, and $\ac_\mu$ is the normal acceleration, i.e., $\ac_\mu \equiv \n^\gamma\cd_\gamma\n_\mu$. The extrinsic curvature is decomposed as
\begin{equation}
\EX \equiv \EC_\mu{}^\mu, \qquad \SH_{\mu\nu} \equiv \EC_{\mu\nu} - \frac{\EX}{3}\h_{\mu\nu},
\end{equation}
where $\SH_{\mu\nu}$ is the shear tensor, and $\EX$ is the expansion factor. Note that we are considering global foliations, and therefore we have zero vorticity (see~\cite{Vitenti2013}). However, since the vorticity is influenced only by vector perturbations, this restriction will not affect the evaluation of the scalar perturbations. The spatial covariant derivative is thus defined by
\begin{equation}
\scd_\alpha M_{\mu_1\dots}^{\indspace\nu_1\dots} = \hp{\cd_\alpha M_{\mu_1\dots}^{\indspace\nu_1\dots}},
\end{equation}
where the projector is defined analogously to the background one defined above, and will shall call spatial any geometrical object which satisfies $\hp{A_\mu} = A_\mu$. The spatial covariant derivative compatible with $\h_{\mu\nu}$ defines the spatial Riemann curvature tensor
\begin{equation}\label{eq:def:SR}
[\scd_\mu\scd_\nu - \scd_\nu\scd_\mu]A_\alpha \equiv \SR_{\mu\nu\alpha}{}^\beta A_\beta,
\end{equation}
where $A_\beta$ is an arbitrary spatial field. For a Friedmann background we have that the extrinsic curvature and the spatial Ricci tensor are diagonal $$\bEC_{\mu\nu} = \frac{\bEX}{3}\bh_{\mu\nu}, \qquad\bSR_{\mu\nu} = 2\bK\bh_{\mu\nu},$$ with the expansion factor and the function $\bK$ being homogeneous, i.e. $\bscd_\mu\bEX = 0 = \bscd_\mu\bK$.

We define the spatial component of the perturbed normal field $\n_\mu$ as $\sn_\mu \equiv \bhp{\dn_\mu}$, where the $\delta$ of a geometrical object ${\bf T}$ is defined as $\delta {\bf T} = {\bf T}- {\bar{\bf T}}$. For scalar perturbations, $\sn_\mu = \bscd_\mu\vms$. Using these definitions, it is easy to show that for any scalar quantity $\mf$, its gradient $\scd_\mu\mf$ at first order can be expressed as
\begin{equation}\label{eq:def:gen:gi}
\delta(\scd_\mu\mf) = \sn_\mu\dot{\bmf} + \bscd_\mu\dmf = \bscd_\mu\left(\vms\dot{\bmf} + \dmf\right),
\end{equation}
where the dot represents the time derivative, i.e., $$\dot{\mf} \equiv \bhp{\lie_\bn\mf} = \bn^\mu\bcd_\mu\mf,$$ and the quantity in parenthesis is automatically gauge invariant. Using the result above, it is easy to calculate the covariant defined gauge invariant variables
\begin{equation*}
\Ced_\mu \equiv \kp\scd_\mu\ed, \quad \Cp_\mu \equiv \kp\scd_\mu\p,\quad \CEX_\mu \equiv \scd_\mu \EX,
\end{equation*}
where $\kp \equiv 8\pi{}G/c^4$, and $G$ is the gravitational constant.

For a barotropic perfect fluid, the Mukhanov-Sasaki equation~\cite{Mukhanov1992} is enough to describe the dynamics. However, since this is a second order time derivative equation, we need two variables to completely describe the state of the system. The Mukhanov-Sasaki variable is defined as
\begin{equation}
\MS = \CSvms - \frac{2\bK\CSI}{\kp(\bed+\bp)},
\end{equation}
where
\begin{equation}
\CSvms \equiv \CS + \frac{\bEX\vms}{3}, \quad \CSI \equiv \CS - \frac{\bEX\dSHs}{3},
\end{equation}
and the scalar shear potential is defined through
\begin{equation}\label{eq:def:FLRW:sCI:decomp}
\begin{split}
\dSH_{\mu\nu} &= \left(\bscd_{(\mu}\bscd_{\nu)}-\frac{\bh_{\mu\nu}\blap}{3}\right)\dSHs \\
&+ \dSHv_{(\nu\scp\mu)} + \dot{\CTD}_\mu{}^\alpha\bh_{\alpha\nu},
\end{split}
\end{equation}
reading,
\begin{equation}\label{eq:FLRW:dSHs}
\dSHs \equiv \left(\BS-\dot{\CSD} + \frac{2}{3}\bEX\CSD\right), \quad \dSHv^{\alpha} \equiv \BV^\alpha + \dot{\CV}^\alpha.
\end{equation}
The perturbation on the expansion factor gives
\begin{equation}\label{eq:FLRW:dEX}
\dEX = \blap\dSHs  + \bEX\A + 3\dot{\CS}.
\end{equation}
The perturbation on the spatial curvature scalar is
\begin{align*}
\dSR &= -4\blapK\CS,
\end{align*}
where we have defined the operator $\blapK \equiv \blap + 3\bK$. Finally, the acceleration of the normal flow defined by $\bn^\mu$ is
\begin{equation}\label{eq:def:dac}
\dac_\mu = -\bscd_\mu\A.
\end{equation}

The kinematic variables above are defined in the background foliation frame and are discussed in details in~\cite{Vitenti2013}. For the barotropic perfect fluid models we have
\begin{equation}
\dot{\MS} = \frac{2\bcs^2\bEX}{3\kp(\bed+\bp)}\bscd^2\CSI.
\end{equation}
Then, recasting the covariant defined gauge invariant variables in terms of $\MS$ and $\CSI$ allows us to map the linearity conditions of K12 directly to our approach. First of all, since in K12 the author only considered spatially flat background models, we put $\bK = 0$.

The kinematic perturbations $\dSR$, $\dSHs$ and $\dEX$ are defined in the background frame. Using the results from Appendix~A of \cite{Vitenti2013}, it is easy to obtain the same variables in the fluid frame,\footnote{The fluid frame is defined by the time-like eigenvector of the energy momentum tensor. The expressions for the perturbations in an arbitrary frame are given by Eqs. (A7), (A14), (A16) and (A30) of~\cite{Vitenti2013} respectively for acceleration, shear, expansion factor and spatial curvature scalar.}
\begin{align}
\dSRf &= \dSR - 4\frac{\bEX\bscd^2\vms}{3} = -4\left(\bscd^2\CSvms + 3\bK\CS\right), \\
\dSHsf &= \dSHs + \vms, \\
\dEXf &= \dEX + \bscd^2\vms = 3\dot{\CS} + \bEX\A + \bscd^2\dSHsf, \\
\dacf_\mu &= \dac_\mu + \bscd_\mu\dot{\vms} = \bscd_\mu(\dot{\vms}-\A),
\end{align}
For null spatial background curvature, the spatial curvature scalar and shear perturbations in the fluid frame are gauge invariant since their background values are null. In order to obtain the gauge invariant variable associated with $\dEXf$, we could calculate $\CEX_\mu$ directly. However, it is easier to use the time-space projection of the Einstein´s equation, i.e.,
\begin{equation}\label{eq:G0i}
\hp{\G_{\n\mu}} = \scd_\nu\SH^\nu{}_\mu - \frac{2}{3}\scd_\mu\EX = 0,
\end{equation}
where the right hand side is null because we are in the fluid frame. Therefore, at first order we have
\begin{equation}
\CEX_\mu = \frac{3}{2}\scd_\nu\SH^\nu{}_\mu \approx \bscd_\mu\bscd^2\dSHsf.
\end{equation}
The acceleration can also be simplified by using the spatial projection of the energy momentum conservation, i.e.,
\begin{equation}
\hp{\cd_\nu\T^\nu{}_\mu} = (\ed + \p)\ac_\mu + \scd_\mu\p = 0.
\end{equation}
Hence, we have
\begin{equation}
\ac_\mu = -\frac{\cs^2\Ced_\mu}{\kp(\ed+\p)},
\end{equation}
where we used that, for a barotropic fluid, $\Cp_\mu = \cs^2\Ced_\mu$ where $\cs^2$ is the fluid speed of sound, i.e., $$\cs^2 \equiv \frac{\del\p}{\del\ed}.$$ From the expressions above, we see that the only variable left to close the set is $\Ced_\mu$. Again, using the time-time projection of the Einstein equation we obtain,
\begin{equation}
\frac{\SR}{2} - \SH^2 + \frac{\EX^2}{3} = \kp\ed.
\end{equation}
Thus, at first order $\Ced_\mu$ can be expressed as
\begin{equation}
\Ced_\mu \approx -2\bscd_\mu\bscd^2\left(\CSvms-\frac{\bEX}{3}\dSHsf\right).
\end{equation}
Finally, in K12 it is also used the following first order variable:\footnote{This expression was originally defined in Eq.~(32) of~\cite{Ellis1989}. This variable is first order only when the background is considered flat.}
\begin{equation}
\RR = -\frac{\SR}{2} + \scd_\mu\ac^\mu - 3\SH^2.
\end{equation}

Using the above results, we can write all the kinematic variables in terms of $\MS$ and $\CSI$ as follows,
\begin{equation}\label{eq:cov:pert:map}
\begin{split}
\dSRf &= -4\bscd^2\MS, \qquad\qquad \dSHsf = \frac{3}{\bEX}\left(\MS-\CSI\right), \\
\CEX_\mu &\approx \frac{3\bscd_\mu\bscd^2}{\bEX}(\MS-\CSI),\quad \ac_\mu \approx \frac{2\bcs^2\bscd_\mu\bscd^2\CSI}{\kp(\bed+\bp)},\\
\Ced_\mu &\approx -2\bscd_\mu\bscd^2\CSI,\qquad\quad \Cp_\mu \approx -2\bcs^2\bscd_\mu\bscd^2\CSI, \\
\RR &\approx 2\bscd^2\left(\MS + \frac{\bcs^2\bscd^2\CSI}{\kp(\bed+\bp)}\right).
\end{split}
\end{equation}

\section{Linearity Conditions}

Using the tools developed in the previous section, we can reanalyze the linearity conditions obtained in K12. Note, however, that the meaning of some expressions in K12 are not precisely defined. For example, the condition $\kcond_2$ given in Eq.~(56) of K12 is written as $$\kcond_2 \equiv \frac{\vert\SH^\nu{}_\mu\Ced_\nu\vert}{\vert\kp(\ed+\p)\CEX_\mu\vert} \ll 1.$$
We understand that $\vert V_\alpha \vert$ is defined as $\vert V_\alpha \vert \equiv \sqrt{\h^{\mu\nu}V_\mu{}V_\nu}$ for any spatial vector field $V_\alpha$. This is possible since the spatial metric is positive definite and, as such, $\h^{\mu\nu}V_\mu{}V_\nu \geq 0$, where the inequality is saturated only when $V_\alpha$ is null.

\subsection{Conditions for Scalar Perturbations}

Using Eqs.~\eqref{eq:cov:pert:map}, we can map the conditions directly in terms of $\MS$ and $\CSI$. However, we still need to deal with the additional spatial derivatives of $\MS$ and $\CSI$. The eigenfunctions $\mode_i$ of the Laplace-Beltrami operator $\bscd^2$,\footnote{The eigenfunctions are defined by $\bscd^2\mode_i = -k^2_i\mode_i$, where $i$ represents the collection of indexes for the mode functions and $k_i$ their eigenvalues.} provide a natural basis for the decomposition of these functions. Thus, decomposing both functions we have $$\CSI = \sum_i\CSI_i\mode_i,\qquad\MS = \sum_i\MS_i\mode_i,$$ where $\sum_i$ represent all sums and integrals necessary. Their Laplacian can be calculated as follows,
\begin{equation*}
\bscd^2\CSI = -\sum_i k^2_i\CSI_i\mode_i,\quad \bscd^2\MS = -\sum_i k^2_i\MS_i\mode_i ,
\end{equation*}
where $k^2_i \equiv \bh_{\mu\nu} k_i^{\mu}k_i^{\nu}$, and $k_i^{\mu}$ is the wave vector.
The first order shear perturbation $\dSH_{\mu\nu}$ has its norm given by
\begin{equation*}
\begin{split}
\vert\dSH_{\mu\nu}\vert^2 &= \sum_{ij}\left((k_{i\mu}k_j{}^\mu)^2-\frac{k_i^2k_j^2}{3}\right)\dSHsf_{i}\dSHsf_{j}\mode_i\mode_j, \\
&\leq \sum_{ij}\left(\frac{2k_i^2k_j^2}{3}\right)\dSHsf_{i}\dSHsf_{j}\mode_i\mode_j = \frac{2}{3}(\blap\dSHsf)^2,
\end{split}
\end{equation*}
where we used the Cauchy-Schwarz inequality to write $(k_{i\mu}k_j{}^\mu)^2 \leq k_i^2k_j^2$. Thus, we can substitute $\dSH_{\mu\nu}$ by $\sqrt{2}\bh_{\mu\nu}\blap\dSHsf/3$ when comparing second and first order terms. Using this result and the map given by Eqs.~\eqref{eq:cov:pert:map}, the condition $\kcond_2$ can be written as
\begin{equation}\label{eq:my:kcond2}
\kcond_2 \leq \frac{2\sqrt{2}\left\vert\bscd^2\left(\MS-\CSI\right)\bscd_\mu\bscd^2\CSI\right\vert}{3\left\vert\bscd_\mu\bscd^2(\MS-\CSI)\kp(\bed+\bp)\right\vert}.
\end{equation}

If we are working with an explicit cut-off in the modes, we define $\mkco$ as the largest eigenvalue. Without a cut-off, we take in the sums above the values of $i$ which give the maximum value of $\vert \mki{i}^2\CSI_i\mode_i\vert$ and $\vert \mki{i}^2\MS_i\mode_i\vert$. From these two values, we choose the one which gives the largest eigenvalue as $\mkco$. Note also that $\dot{\mki{i}^2} = -2\bEX\mki{i}^2/3$, and for this reason we define $$\mkcoz^2 \equiv \x^{-2}\mkco^2,\quad \dmkcoz = 0,$$ where $\x = a_0/a$, $a$ is the background scale factor, $3\dot{a}/a = \bEX$, and $a_0$ is the scale factor at a fixed time. We say that a mode $i$ is in super Hubble evolution when the potential that drives its evolution becomes larger than $\mki{i}^2$. Since the eigenvalues are arranged in a monotonically increasing sequence we say that a perturbation is in a super Hubble evolution when the largest mode $\mkco$ enters in super Hubble evolution.


In order to evaluate the super Hubble behavior of Eq.~\eqref{eq:my:kcond2}, and recalling that the square spatial norm of any vector field $V_\mu$, $\vert V_\mu\vert^2 = V_{\mu}V_\nu\bh^{\mu\nu}$, contains a $\bh^{\mu\nu}$ which evolves proportionally to $x^2$, we write $\kcond_2$ as
\begin{equation}
\kcond_2 \leq f_2\frac{\left\vert\bscd_\mu\bscd^2\CSI\right\vert}{\left\vert\kp\x(\bed+\bp)\right\vert}, \quad f_2 \equiv \frac{2\sqrt{2}\left\vert\x\bscd^2\left(\MS-\CSI\right)\right\vert}{3\left\vert\bscd_\mu\bscd^2(\MS-\CSI)\right\vert},
\end{equation}
where the function $f_2$ in super Hubble evolution is constant in time. From Eqs.~(10), (14), (18) and (40) in VPN12, we have the evolution of $\CSI$ and $\MS$ in super Hubble scales for a single fluid, i.e.,
\begin{equation}\label{eq:CSI:MS:growth}
\CSI = \CSI_{0}\x^{(5+3\bw)/2},\qquad \MS = \MS_{0}\x^{3(1-\bw)/2},
\end{equation}
where $\bw $ is the equation of state parameter of the fluid ($\bw = \bp/\bed$), and the subscript ${}_{0}$ refers to the function calculated at that same instant as $a_0$. Hence, for these models we obtain
\begin{equation}
\label{27}
\kcond_2 \leq f_2\frac{\left\vert(\bscd_\mu\blap\CSI)_0\right\vert}{\left\vert\kp(\bed_0+\bp_0)\right\vert}\x^{3(1-\bw)/2}.
\end{equation}

For radiation we have $\kcond_2 \propto \x \propto a^{-1}$, the same result as K12. This is exactly what we have already obtained in VPN12. There we have also shown that if $\MS \ll 1$, then it would be enough to have Eq.~(\ref{27}) and all other linear conditions satisfied in order to keep the linear regime valid. For radiation, we have $\MS = \MS_0\x$. Hence, it is a crucial step to estimate whether $\MS \ll 1$ cease to be valid in bouncing models in order to check the validity of the linear regime. The maximum value of $\MS$ happens at the bounce. In that case, from Eqs.~(12, 13, 18, 41) of VPN12 and from Ref.~\cite{Peter2007}, one obtains that the modes of $\MS$ during the super Hubble evolution have their power spectrum given by $$\Delta_{\MS_ib}^2 = n^2 \left(\frac{\lp}{\HR}\right)^2 \mki{i}^{n_s-1} x_b^{3(1-\bw)},$$ where $n$ is a mode independent constant, $\lp \equiv \sqrt{\kp\hbar{}c}$ is the Planck length, $\HR = c/H_0$ is the present Hubble radius, $n_s$ is the scalar spectral index, and $\mki{i}$ is in Hubble radius units.

The constant $n$ depends on the equation of state parameter $w_0$ of the matter content which is dominating the background when the scales of cosmological interest today are getting bigger than the curvature scale of the universe in the contracting phase. In order to have $n_s\approx 1$, one has to have $|w_0|\approx 0$, and the matter content should be like cold dark matter. If this matter content is described by a hydrodynamical fluid, then $n \approx w_0^{3(w_0-1)/[4(1+3w_0)]} \approx w_0^{-3/4} \gg 1$.\footnote{For an asymmetric bounce, the constant $n$ can also be large if the contracting Hubble rate is larger then the expanding one at the same value of the scale factor, or if there is particle creation near the bounce.}

The variance of $\MS$ at a given point at the bounce can be evaluated from the above expression, and it is given by
\begin{equation}
\begin{split}
&\left\langle\MS^2\right\rangle_b = \int_{\mki{i_\text{min}}}^{\mkco}\Delta_{\MS_ib}^2\frac{d\mki{i}}{\mki{i}}, \\
&\approx \frac{n^2}{\vert n_s-1\vert}\left(\frac{\lp}{\HR}\right)^2 x_b^{3(1-\bw)}\left\vert\mkco^{n_s-1} - \mki{i_\text{min}}^{n_s-1}\right\vert, \\
&\approx \frac{n^2}{\vert n_s-1\vert}x_b^{3(1-\bw)}\left(\frac{\lp}{\HR}\right)^2,
\end{split}
\end{equation}
where $\mki{i_\text{min}}$ represents the largest observable wavelength. In the last line above we considered $n_s \approx 1$ but smaller, as observations indicate, and $\mki{i_\text{min}} \approx 1$. Then we expect that $$\MS_b \approx \frac{n\x_b^{3(1-\bw)/2}}{\sqrt{\vert n_s-1\vert}}\frac{\lp}{\HR} \approx \frac{n}{\sqrt{\vert n_s-1\vert}} 10^{-60}\x_b^{3(1-\bw)/2},$$ within some standard deviations. As $\MS$ remains constant afterwards, $\MS_b$ is its maximum value then, and it is constrained by observations to be $\MS_o \approx 10^{-5} \approx \MS_b$. Hence, if this condition is satisfied, we are guaranteed that perturbations remain linear all along. 

We must now see what constraints on the bouncing background this condition implies. For the case of a radiation dominated bounce, one has that 
\begin{equation}
\begin{split}
\MS_b &\approx \frac{n}{\sqrt{\vert n_s-1\vert}} \frac{x_b \lp}{\HR} \approx 10^{-5} \\
&\Rightarrow\quad x_b \equiv \frac{a_0}{a_b} \approx 10^{55} \frac{\sqrt{\vert n_s-1\vert}}{n},
\end{split}
\end{equation}
where $a_b$ is the value of the scale factor at the bounce.
Hence we have $a_0/a_b < 10^{54}/n$, where we have assumed the observed value $\vert n_s-1\vert \approx 10^{-2}$. As we have seen above, for a dust dominated contraction $n \gg 1$ ($|w_0|\approx 0$). Hence, depending on the value of $w_0$, one can obtain the reasonable interval $10^{12} < a_0/a_b < 10^{33}$, where the bounce happens at energy scales above nucleosynthesis and below Planck energy scales. 

The discussion above shows that the amplitude of $\MS$ is controlled even for very deep bounces. In fact, contrary to usual intuition, if it were not for the factor $n$ above, even a bounce at Planck energy scale would lead to perturbation amplitudes much less then the ones observed from the CMB anisotropies, $\MS_o \approx 10^{-5}$. Hence, all linearity conditions will be largely satisfied at the bounce itself and for any $x = a_0/a$ smaller than $x_b$. Such discussion is missing in K12, where the value of $\kcond_2$ is not estimated at any time, only the growth rate is obtained for the super Hubble scales. Then, from this result it is mistakenly concluded, in the abstract of K12, that the linearity conditions will not be satisfied near the bounce phase.

Note also that with the scale factor behavior of $\SH^S$ given in Eq.~(94) of K12, $$\SH^S = \Sigma_S^{(1)}a^{-1} + \Sigma_S^{(2)}a^{2},$$  one would not have obtained the correct scale factor dependence of $\kcond_2$. In fact, Eq.~(94) of K12 is not correct because it is in contradiction with the constraint given in Eq.~\eqref{eq:G0i} [Eq.~(53) in K12], which implies that $$\SH^S = \frac{Z^{(1)}}{k}a^{-3} + \frac{Z^{(2)}}{k}.$$ However, as K12 considered all kinds of perturbations (scalar, vector and tensor) when calculating the shear, and, since the tensor part grows as $a^{-3}$, he finally obtained the correct result for $\kcond_2$ in the radiation case.

For the dust fluid, we calculated above that $\kcond_2 \propto a^{-3/2}$, while in K12 it was obtained that $\kcond_2 \propto a^{-3/2}\ln(a)$. However, it is easy to see that K12 is not correct: as in the radiation case, the scale factor dependence of $\SH^S$ was also mistakenly calculated in K12, $$\SH^S = \Sigma_S^{(1)}a^{-3}\ln(a) + \Sigma_S^{(2)}a^{1/2},$$ which is in sharp contradiction with the constraint $$\SH^S = \frac{Z^{(1)}}{\mki{}}a^{-3} + \frac{Z^{(2)}}{\mki{}}a^{-1/2},$$ which would give the correct growth rate for $\kcond_2 \propto a^{-3/2}$ for the dust fluid.

Note also that, for both matter contents, the growth rate of the shear is $a^{-3}$. This is exactly what one obtains from Eq.~\eqref{eq:cov:pert:map} and Eq.~\eqref{eq:CSI:MS:growth}, which gives $\dSHsf \propto \x$ and, therefore, $\dSH_{\mu\nu} \propto \dSH_{\mu\nu}(a_0)a^{-3}$.

This growth rate for the shear tensor is commonly called anisotropy problem in bouncing models (see, for instance,~\cite{Bozza2009} and references therein). The problem appears when one confront the growth rate of $\EX^2 \propto \x^{3(1+\bw)}$ and $\SH^2 \propto \x^6$. Hence, for any contracting phase, the shear grows faster if the fluid has equation of state less than that of a stiff matter, i.e., $\bw < 1$.

However, when the shear arises from perturbations generated by quantum vacuum fluctuations, this problem does not appears. In this case, $\SH^2$ is a second order variable, while $\EX^2$ is a zero order one. Hence, using Eqs.~(\ref{eq:cov:pert:map}) and (\ref{eq:CSI:MS:growth}), one obtains that (for the case of radiation domination at the bounce, which implies $\EX^2 \propto x^4$) $$\frac{\SH^2}{\EX^2} \approx \left(\frac{\lp}{\HR}\right)^2\x^2 \approx 10^{-120}\x^2, \quad \left.\frac{\SH^2}{\EX^2}\right\vert_b \leq 10^{-8}.$$ Hence, with respect to the shear, a Friedmann background plus initial quantum vacuum perturbations can accommodate a wide class of bouncing models without appealing to any fine tuning.

The next condition $\kcond_3$ can be written as $$\kcond_3 \equiv \frac{2\vert\RR\ac_\mu\vert}{\vert\Ced_\mu\vert} = \left\vert\frac{4\bcs^2\bscd^2}{\kp(\bed+\bp)}\left(\MS + \frac{\bcs^2\bscd^2\CSI}{\kp(\bed+\bp)}\right)\right\vert \ll 1.$$ It provides a much less stringent constraint on the contracting phase. Using Eq.~\eqref{eq:CSI:MS:growth}, we obtain $$\kcond_3 = \left\vert\left[\frac{4\bcs^2\blap}{\kp(\bed+\bp)}\left(\MS + \frac{\bcs^2\blap\CSI}{\kp(\bed+\bp)}\right)\right]_0\right\vert \x^{(1-9\bw)/2}.$$ In the radiation case, we have $\kcond_3 \propto a$. These results do not agree with that of K12. This mistake of K12 happens because in Eq.~(106) of K12 the second order term in $\RR$ was used to calculate its growth. However, this is the same as comparing a third order term $\SH^2\ac_\mu$ with a first order one $\Ced_\mu$, which leads naturally to a faster growth rate. If one had used the correct first order term $\bar{A}a^{-3}$ (in the notation of K12), the correct growth rate of $\kcond_3 \propto a$ would have been obtained.

The conditions $\kcond_5$ and $\kcond_7$ are closely related,\footnote{Note that in Eqs.~(110) and (112) of K12 there are typos, $\Omega^b{}_a$ should read $\Sigma^b{}_a$ in the expression for $\kcond_7$ and $\kcondac_7$.} i.e.,
\begin{align*}
\kcond_5 &\equiv \frac{\vert4\scd_\mu\SH^2\vert}{\vert\Ced_\mu\vert} \leq \frac{12}{\bEX^2}\frac{\vert\bscd_\mu\bscd^2(\MS-\CSI)\bscd^2(\MS-\CSI)\vert}{\vert\bscd_\mu\bscd^2\CSI\vert}, \\
\kcond_7 &\equiv \frac{\left\vert2\SH_{\mu}{}^\nu\CEX_\nu\right\vert}{\left\vert\Ced_\mu\right\vert} \leq \frac{3\sqrt{2}}{\bEX^2}\frac{\vert\bscd_\mu\bscd^2(\MS-\CSI)\bscd^2(\MS-\CSI)\vert}{\vert\bscd_\mu\bscd^2\CSI\vert}.
\end{align*}
For the two cases analyzed here, dust and radiation, $\CSI$ grows faster than $\MS$ and $\bEX^2\propto\bed$. In this case, both constraints are proportional to $\kcond_2$, i.e., $$\kcond_5\propto\kcond_7\propto \kcond_2 \propto \x^{3(1-\bw)/2}.$$ For this reason, these constraints do not add new restrictions to the gauge invariant variables.

The other set of constraints in K12 were obtained by comparing the second order with $\scd_\mu\scd_\nu\ac^\nu$ instead of $\Ced_\mu/2$. Since these two quantities are proportional to each other, we can write
\begin{equation}
\kcondac_a = \frac{\vert\Ced_\mu\vert}{2\vert\scd_\mu\scd_\nu\ac^\nu\vert}\kcond_a,
\end{equation}
for $a = 3,5,7$. This add an additional factor of $$\left\vert\frac{\kp(\bed+\bp)\bscd_\mu\blap\CSI}{2\bcs^2\bscd_\mu\blap\blap\CSI}\right\vert = \left\vert\left[\frac{\kp(\bed+\bp)\bscd_\mu\blap\CSI}{2\bcs^2\bscd_\mu\blap\blap\CSI}\right]_0\right\vert \x^{1+3\bw}.$$ This implies an additional factor in the constraint proportional to $a^{-2}$ and $a^{-1}$ for radiation and dust, respectively. These conditions grow faster than the first set. Notwithstanding, only one set must be fulfilled, since one must compare the second order terms with the largest first order ones.

The calculations above showed that the original conclusions of VPN12 are correct. The value of $\MS$ (or the constant mode of $\CSI$) do not get larger than one near the bounce, and there is a gauge in which all perturbations remain small.

\subsection{Conditions for Vector and Tensor Perturbations}

In K12 the author also considered conditions mixing scalar, vector and tensor perturbations. These constraints are originated by the presence of vector and tensor perturbations in both the shear $\SH_{\mu\nu}$ and the vorticity. For the tensor perturbations, it is easy to show that it also contributes to the shear with a factor which grows with $a^{-3}$. When considering the vacuum initial conditions for both scalar and tensor perturbations, the tensor contribution to the shear will also be of first order. In addition, for some class of models it is shown that the tensor amplitude is indeed smaller than the scalars' (see, for instance~\cite{Peter2007,Bessada2012}), hence, even if we consider its contribution to the shear, the constraints remain valid.

On the other hand, the vector perturbations have a very simple evolution. However, this fact makes the quantization and the imposition of similar initial conditions for the vector modes a non straightforward procedure. Without a theory of initial conditions for them, the growth rate for the vector perturbations it is not of much use. Notwithstanding, one can heuristically assume that the initial conditions for the vector perturbations are also of the same order of the scalar and tensor perturbations. Then, in this case, the growth rate are not larger than that of $\MS$ and, therefore, one can still use the value of $\MS$ as a measure of the validity of the perturbative series.

\section{Concluding Remarks}

We have shown in this comment that the constraints obtained by the comparison between second order and first order terms in the K12 paper reduce to those obtained in VPN12. The difference in the calculations arises from the complication introduced by the presence of spatial derivatives in the K12 constraints, which we have treated. The idea of this comparison is interesting as a natural extension of our work VPN12. However, the conclusions of K12 are not new and some of them are not correct because estimates of the amplitudes of the perturbations are missing in K12, as we discussed throught this paper. Hence, the conclusions of VPN12 are correct, namely, that cosmological perturbations of quantum mechanical origin in the contracting phase of a bouncing model can remain in the linear regime for a wide interval of energy scales of the bounce, ranging from the nucleosynthesis energy scale up to some few orders of magnitude below the Planck energy.

\section*{ACKNOWLEDGMENTS}

We would like to thank CNPq of Brazil for financial support.  We also would like to thank ``Pequeno Semin\'{a}rio'' of CBPF's Cosmology Group for useful discussions, comments and suggestions.

\end{document}